\documentclass[aps,prb,twocolumn,showpacs,preprintnumbers,amsmath,amssymb,superscriptaddress]{revtex4}%

\usepackage{graphicx}
\usepackage{dcolumn}
\usepackage{bm}
\usepackage{color}

\begin{document}

\preprint{preprint(\today)}

\title{Muon-spin rotation measurements of the magnetic penetration depth in the Fe-based superconductor  Ba$_{1-x}$Rb$_ {x}$Fe$_{2}$As$_{2}$}

\author{Z.~Guguchia}
\email{zurabgug@physik.uzh.ch} \affiliation{Physik-Institut der
Universit\"{a}t Z\"{u}rich, Winterthurerstrasse 190, CH-8057
Z\"{u}rich, Switzerland}

\author{Z.~Shermadini}
\affiliation{Laboratory for Muon Spin Spectroscopy, Paul Scherrer Institute, CH-5232
Villigen PSI, Switzerland}

\author{A.~Amato}
\affiliation{Laboratory for Muon Spin Spectroscopy, Paul Scherrer Institute, CH-5232
Villigen PSI, Switzerland}

\author{A.~Maisuradze}
\affiliation{Physik-Institut der Universit\"{a}t Z\"{u}rich,
Winterthurerstrasse 190, CH-8057 Z\"{u}rich, Switzerland}
\affiliation{Laboratory for Muon Spin Spectroscopy, Paul Scherrer Institute, CH-5232
Villigen PSI, Switzerland}

\author{A.~Shengelaya}
\affiliation{Department of Physics, Tbilisi State University,
Chavchavadze 3, GE-0128 Tbilisi, Georgia}

\author{Z.~Bukowski}
\affiliation{Laboratory for Solid State Physics, ETH Z\"urich, CH-8093 Z\"{u}rich, Switzerland}
\affiliation{Institute of Low Temperature and Structure Research, Polish Academy of Sciences,
50-422 Wroclaw, Poland}

\author{H.~Luetkens}
\affiliation{Laboratory for Muon Spin Spectroscopy, Paul Scherrer Institute, CH-5232
Villigen PSI, Switzerland}

\author{R.~Khasanov}
\affiliation{Laboratory for Muon Spin Spectroscopy, Paul Scherrer Institute, CH-5232
Villigen PSI, Switzerland}

\author{J.~Karpinski}
\affiliation{Laboratory for Solid State Physics, ETH Z\"urich,
CH-8093 Z\"{u}rich, Switzerland}

\author{H.~Keller}
\affiliation{Physik-Institut der Universit\"{a}t Z\"{u}rich,
Winterthurerstrasse 190, CH-8057 Z\"{u}rich, Switzerland}

\begin{abstract}
Measurements of the magnetic penetration depth ${\lambda}$ in the Fe-based superconductor
Ba$_{1-x}$Rb$_{x}$Fe$_{2}$As$_{2}$ ($x=0.3$, 0.35, 0.4) were carried out using 
the muon-spin rotation (${\mu}$SR) technique. The temperature dependence of
${\lambda}$ is well described by a two-gap $s$+$s$-wave 
scenario with a small gap ${\Delta}$$_{1}$ ${\approx}$ 1 - 3 meV and
a large gap ${\Delta}$$_{2}$ ${\approx}$ 7 - 9 meV.  
By combining the present data with those previously
obtained for RbFe$_{2}$As$_{2}$ a decrease of the BCS ratio
2${\Delta}$$_{2}$/$k_{\rm B}$$T_{\rm c}$ with increasing Rb content $x$ is observed. On the other hand, the BCS ratio 2${\Delta}$$_{1}$/$k_{\rm B}$$T_{\rm c}$ is almost independent
of $x$. In addition, the contribution of ${\Delta}_{1}$ to the superfluid density is found to increase with $x$. 
These results are discussed in the light of the suppression of interband processes upon hole doping.

\end{abstract}


\maketitle

\section{Introduction}
The discovery\cite{Kamihara08} of superconductivity in iron oxypnictide
LaFeAsO$_{1-x}$F$_{x}$ has generated 
a great interest in the
phenomenon of high temperature superconductivity.
The basic units
responsible for superconductivity are the fluorite type
[Fe$_{2}$$Pn$$_{2}$] layers where $Pn$ is a pnictogen element (P, As, Sb,
and Bi). These layers are separated by spacer layers
which play the role of a charge reservoir. In the fluorite
type layers the Fe atoms are surrounded by four pnictogen
atoms forming a tetrahedron. The first class of iron-based superconductors
studied has the ZrCuSiAs structure (1111 compounds),
where the spacer layer [$Ln_{2}$O$_{2}$] has the ''antifluoride''
or Pb$_{2}$O$_{2}$ structure. With $Ln$ = Sm a critical temperature
higher than 55 K was observed.\cite{ZAREN}

 Superconductivity with $T_{\rm c}$ = 38 K was also found in
the ternary systems $A$Fe$_{2}$As$_{2}$ \cite{Rotter,Bukowski} (122 compounds) adopting the
tetragonal ThCr$_{2}$Si$_{2}$ structure. In this structure the spacer layer is
provided by an alkali earth element $A$ = Ca, Sr, or Ba.
Doping is realized by the substitution of $A$ by an alkali
metal such as K, Cs or Rb. Several disconnected Fermi-surface
sheets contribute to superconductivity as revealed by angle-resolved photoemission
spectroscopy (ARPES).\cite{Evtushinsky,Ding,Zabolotnyy} Moreover, indications of multi-gap 
superconductivity in the system Ba$_{1-x}$K$_{x}$Fe$_{2}$As$_{2}$ 
were obtained from the temperature dependence of the magnetic penetration depth ${\lambda}$
by means of muon-spin rotation (${\mu}$SR) \cite{Khasanov} and ARPES.\cite{Evtushinsky}
The magnetic penetration depth is one of the fundamental parameters of a superconductor since it is closely related to the 
density of the superconducting carriers $n_{s}$ and their effective mass $m^*$ via the relation 1/${\lambda}^{2} \propto n_s/m^*$. 
The temperature dependence of ${\lambda}$ reflects the topology of the superconducting gap occurring in the density of states of the superconducting ground state. 
The ${\mu}$SR technique provides a powerful tool to measure ${\lambda}$
in type II superconductors.\cite{Sonier}

 As demonstrated in previous works,\cite{Bukowski,Shermadini} the value of 
$T_{\rm c}$ for hole-doped Ba$_{1-x}$Rb$_{x}$Fe$_{2}$As$_{2}$
decreases monotonically upon increasing the Rb content $x$ in the over-doped region. 
However, in contrast to the over-doped cuprates, $T_{\rm c}$ remains finite even at the highest doping level $x=1$ with $T_{\rm c}=2.52$~K.\cite{Bukowski} 
A detailed study of the doping dependence of $T_{\rm c}$ may help to clarify the origin of high-$T_{\rm c}$ superconductivity in these iron-based systems. 
It is thus of importance to investigate the 
superconducting properties of optimally Ba$_{1-x}$Rb$_{x}$Fe$_{2}$As$_{2}$  and compare 
the results with those obtained for RbFe$_{2}$As$_{2}$.\cite{Shermadini}

 In this paper, we report on ${\mu}$SR studies of 
the temperature and field dependence of the magnetic penetration depth of optimally doped Ba$_{1-x}$Rb$_{x}$Fe$_{2}$As$_{2}$ ($x=0.3$, 0.35, 0.4). We compare the present data with the previous results of overdoped RbFe$_{2}$As$_{2}$  \cite{Shermadini}
and discuss the combined results
in the light of the suppression of interband processes upon hole doping.

\section{EXPERIMENTAL DETAILS }
 Polycrystalline samples of Ba$_{1-x}$Rb$_{x}$Fe$_{2}$As$_{2}$ were prepared in evacuated quartz ampoules 
by a solid state reaction method. Fe$_{2}$As, BaAs, and RbAs were obtained by 
reacting high purity As (99.999 $\%$), Fe (99.9$\%$), Ba (99.9$\%$) and Rb (99.95$\%$) 
at 800 $^{\circ}$C, 650 $^{\circ}$C and 500~$^{\circ}$C, respectively.
Using stoichiometric amounts of BaAs or RbAs and Fe$_{2}$As the terminal compounds BaFe$_{2}$As$_{2}$ 
and RbFe$_{2}$As$_{2}$ were synthesized at 
950 $^{\circ}$C and 650 $^{\circ}$C, respectively. 
Finally, the samples of Ba$_{1-x}$Rb$_{x}$Fe$_{2}$As$_{2}$ with $x$ = 0.3, 0.35, 0.4 were prepared from appropriate 
amounts of single-phase BaFe$_{2}$As$_{2}$ and RbFe$_{2}$As$_{2}$. The components were mixed, pressed into pellets, 
placed into alumina crucibles and annealed for 100 hours at 650 $^{\circ}$C with one 
intermittent grinding. Powder X-ray 
diffraction analysis revealed that the synthesized samples 
are single phase materials. Zero-field (ZF) and transverse-field (TF) ${\mu}$SR 
experiments were performed at the ${\pi}$M3 beamline of the Paul Scherrer
Institute (Villigen, Switzerland), using the general purpose instrument (GPS). The sample was mounted inside of a
gas-flow $^4$He cryostat on a sample holder with a standard veto setup providing essentially a 
low-background $\mu$SR signal. All TF experiments were carried out after a field-cooling procedure.


\section{RESULTS AND DISCUSSION}
\begin{figure}[b!]
\includegraphics[width=0.98\linewidth]{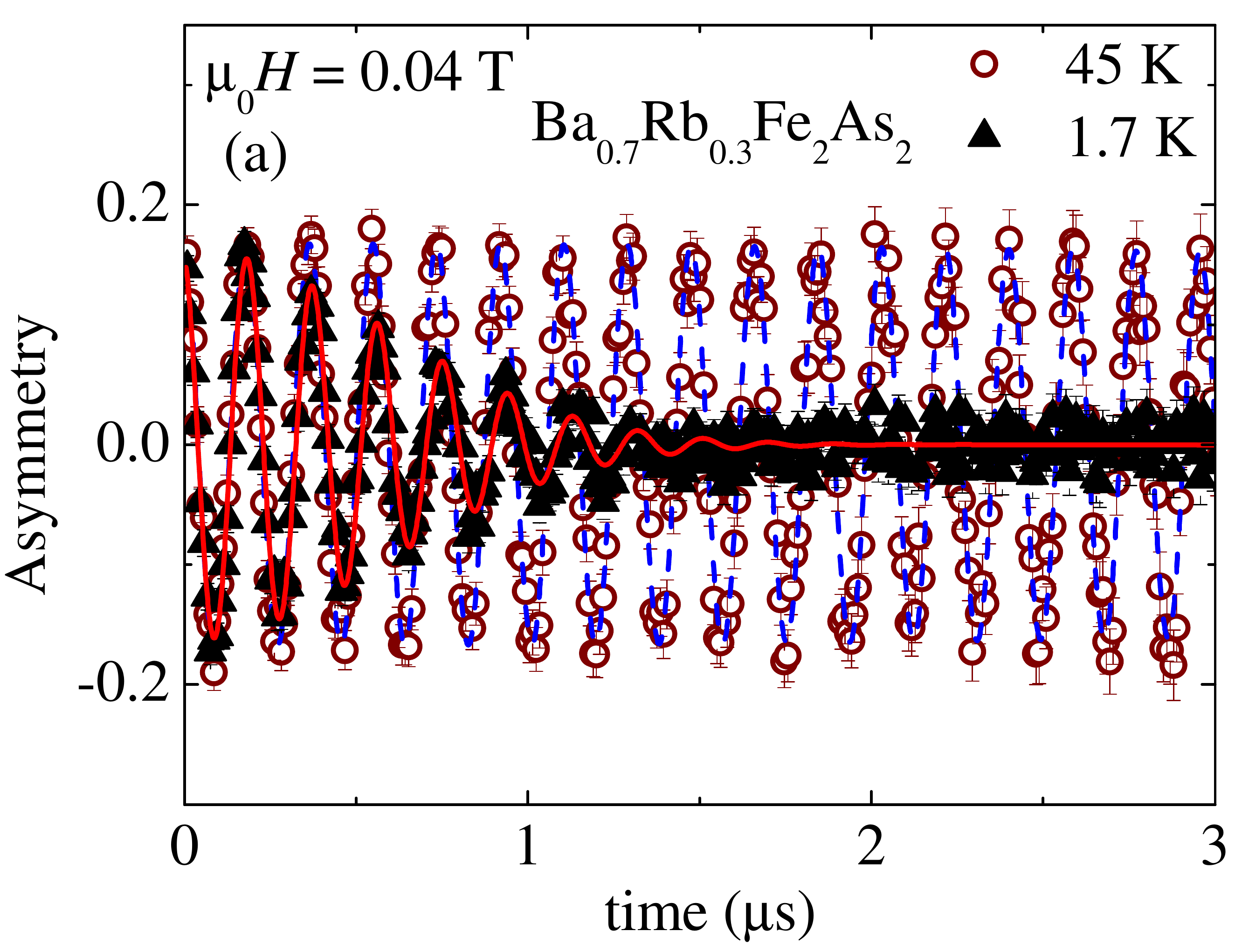}
\includegraphics[width=0.98\linewidth]{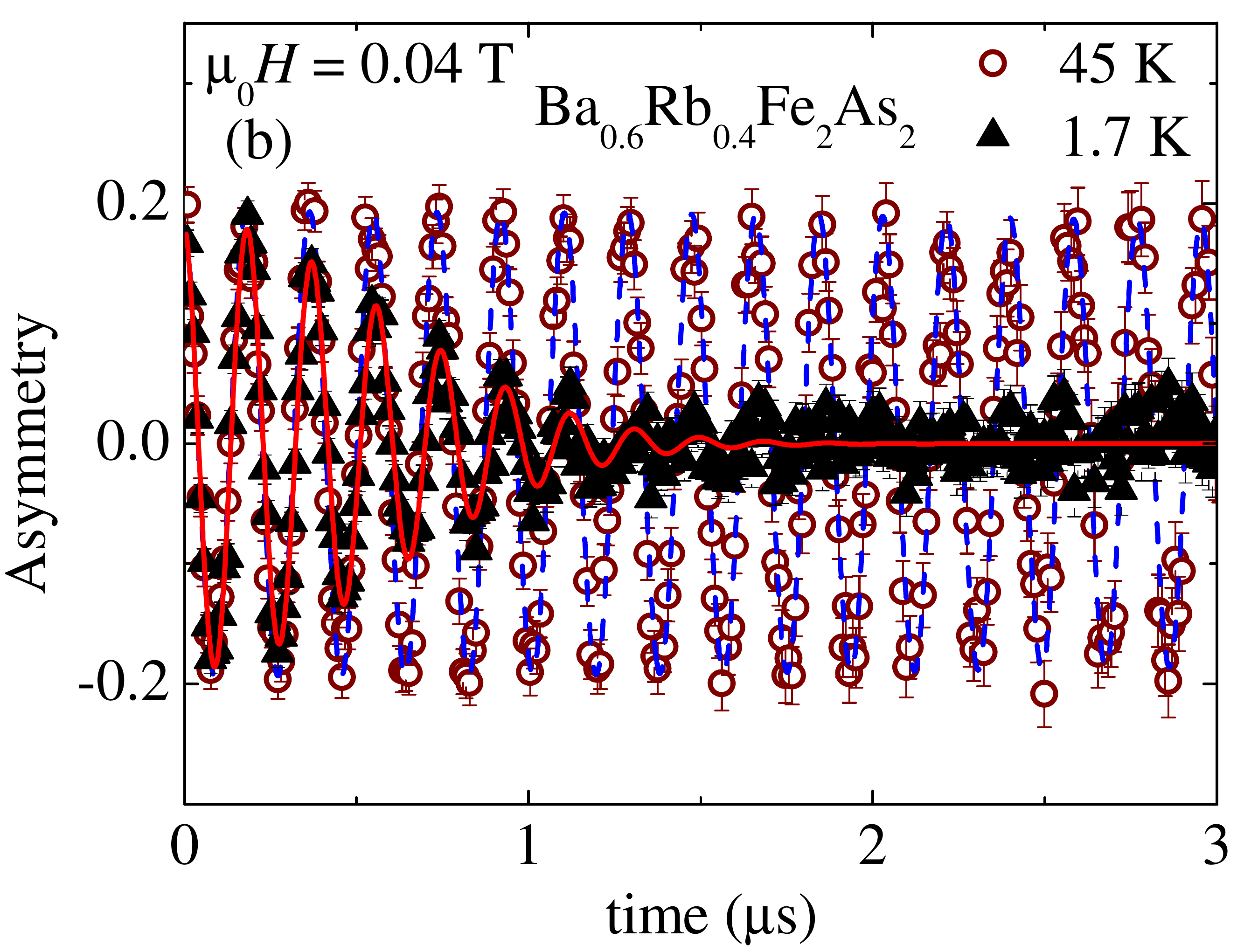}
\caption{ (Color online) Transverse-field (TF) ${\mu}$SR time spectra obtained in ${\mu}$$_{\rm 0}$$H$ = 0.04 T
above and below $T_{\rm c}$ (after field cooling the sample from above $T_{\rm c}$): 
(a) Ba$_{0.7}$Rb$_{0.3}$Fe$_{2}$As$_{2}$ and (b) Ba$_{0.6}$Rb$_{0.4}$Fe$_{2}$As$_{2}$. The solid and the dashed lines represent fits to the data by means of Eq.~(\ref{eq1}).}
\label{fig1}
\end{figure}
\begin{figure}[b!]
\includegraphics[width=0.98\linewidth]{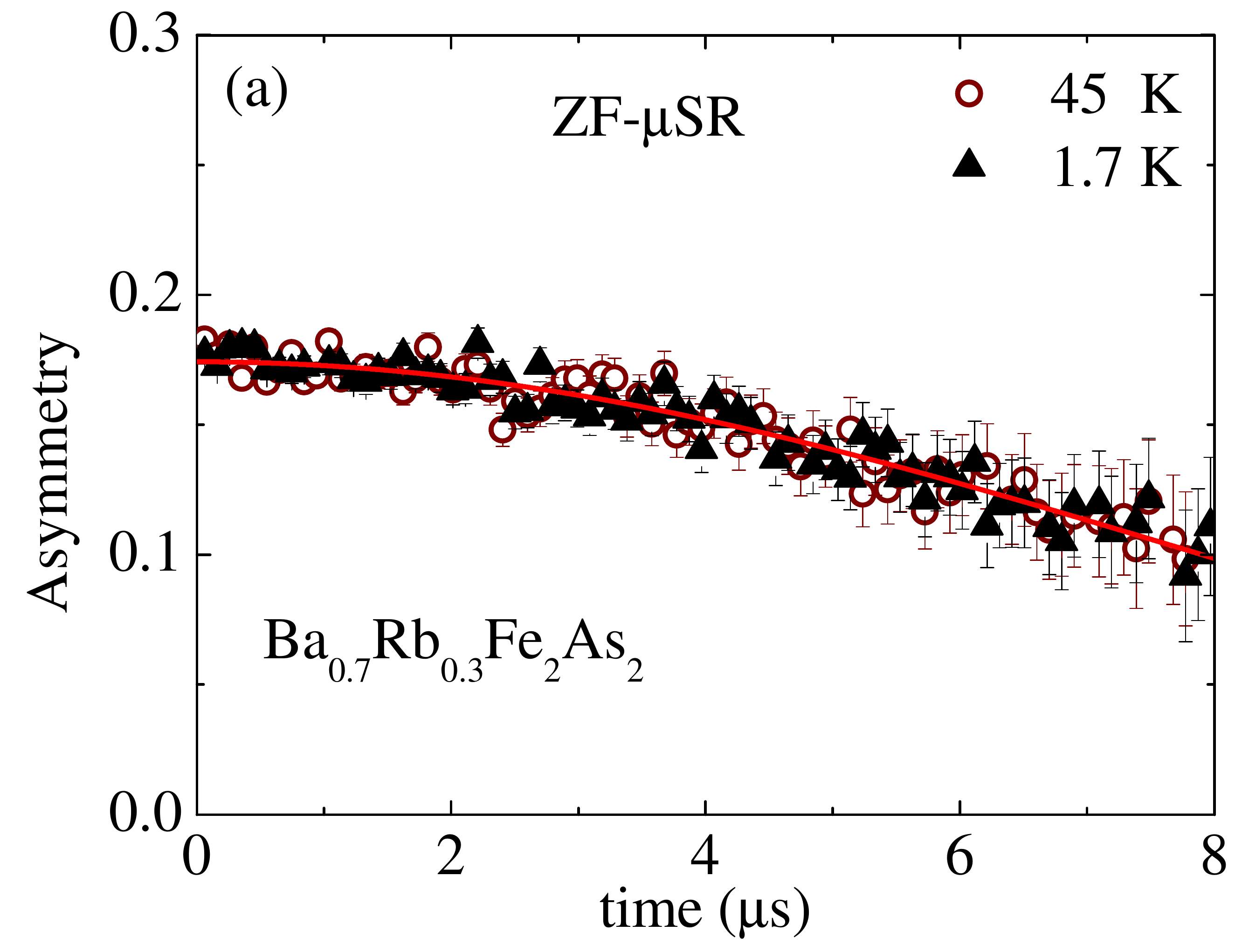}
\includegraphics[width=0.98\linewidth]{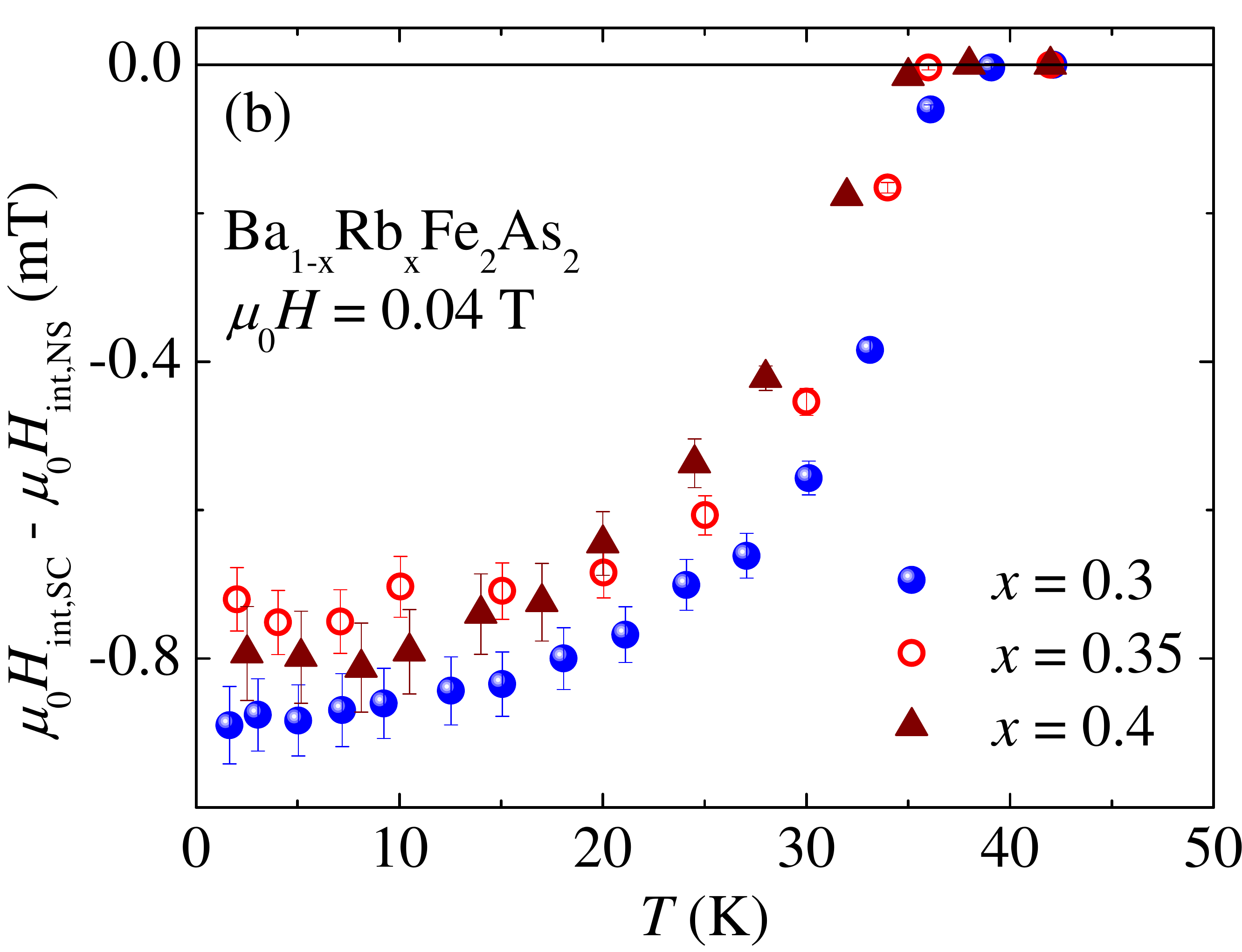}
\caption{(Color online) (a) ZF-${\mu}$SR time spectra for Ba$_{0.7}$Rb$_{0.3}$Fe$_{2}$As$_{2}$ 
recorded above and below $T_{\rm c}$. The line represents the fit to the data
of a standard Kubo-Toyabe depolarization function,\cite{Kubo} (b) Temperature dependence of the difference
between the internal field ${\mu}_{\rm 0}$$H_{\rm int,SC}$ measured in the SC state 
and the one measured in the normal state ${\mu}_{\rm 0}$$H_{\rm int,NS}$ at $T$ = 42~K.}
\label{fig7}
\end{figure}
 
Figures \ref{fig1}a and \ref{fig1}b exhibit the transverse-field (TF) muon-time spectra for Ba$_{1-x}$Rb$_{x}$Fe$_{2}$As$_{2}$ ($x=0.3$, 0.4) 
measured in an applied magnetic field of ${\mu}$$_{\rm 0}$$H$ = 0.04~T above (45 K) and below (1.7 K) the superconducting (SC) transition temperature $T_{\rm c}$. Above $T_{\rm c}$ the oscillations show a small relaxation
due to the random local fields from the nuclear magnetic moments.
Below $T_{\rm c}$ the relaxation rate strongly increases due to the presence of a
nonuniform local field distribution as a result of the formation of a flux-line lattice (FLL) in the SC state.
It is well known that undoped BaFe$_{2}$As$_{2}$ is not superconducting
at ambient pressure and undergoes a spin-density wave (SDW) transition of the Fe-moments far above $T_{\rm c}$.\cite{Huang} The SC state can be achieved either
under pressure \cite{Torikachvili,Miclea} or by appropriate
charge carrier doping \cite{Zhao} of the parent compounds, leading to a suppression of the SDW state.
Magnetism, if present in the samples, may enhance the muon depolarization rate
and falsify the interpretation of the TF-${\mu}$SR results.
Therefore, we have carried out ZF-${\mu}$SR experiments above and below $T_{\rm c}$ to search for
magnetism (static or fluctuating) in Ba$_{1-x}$Rb$_{x}$Fe$_{2}$As$_{2}$ ($x = 0.3$, 0.35, 0.4).
As shown in Fig.~2a no sign of 
either static or fluctuating magnetism could be detected in 
ZF time spectra down to 1.7 K. Moreover, the ZF relaxation rate is small and changes very 
little between 45 K and 1.7 K. The spectra are well described by a standard Kubo-Toyabe
depolarization function,\cite{Kubo} reflecting the field distribution at the muon site 
created by the nuclear moments.

\begin{figure}[t!]
\includegraphics[width=0.98\linewidth]{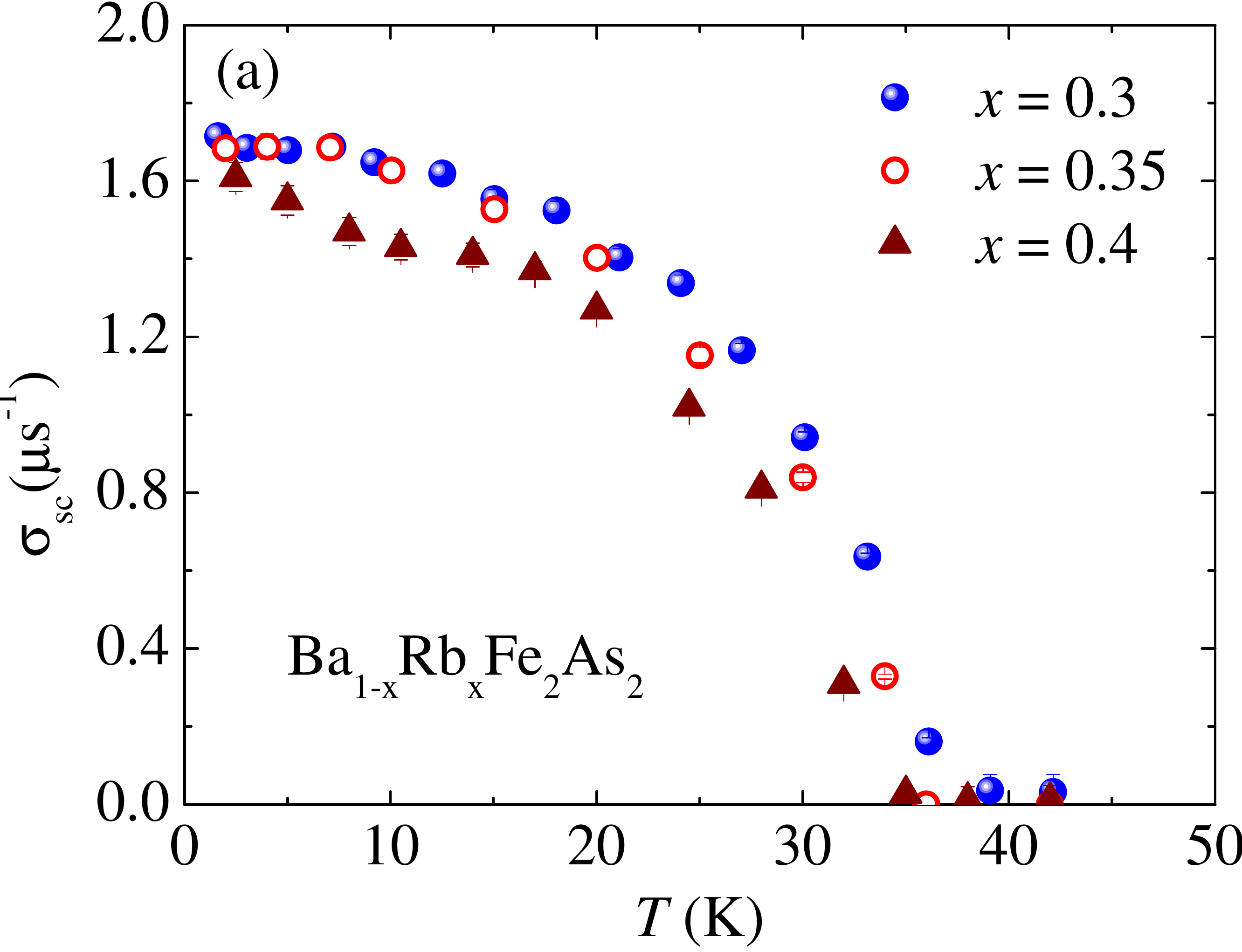}
\includegraphics[width=0.98\linewidth]{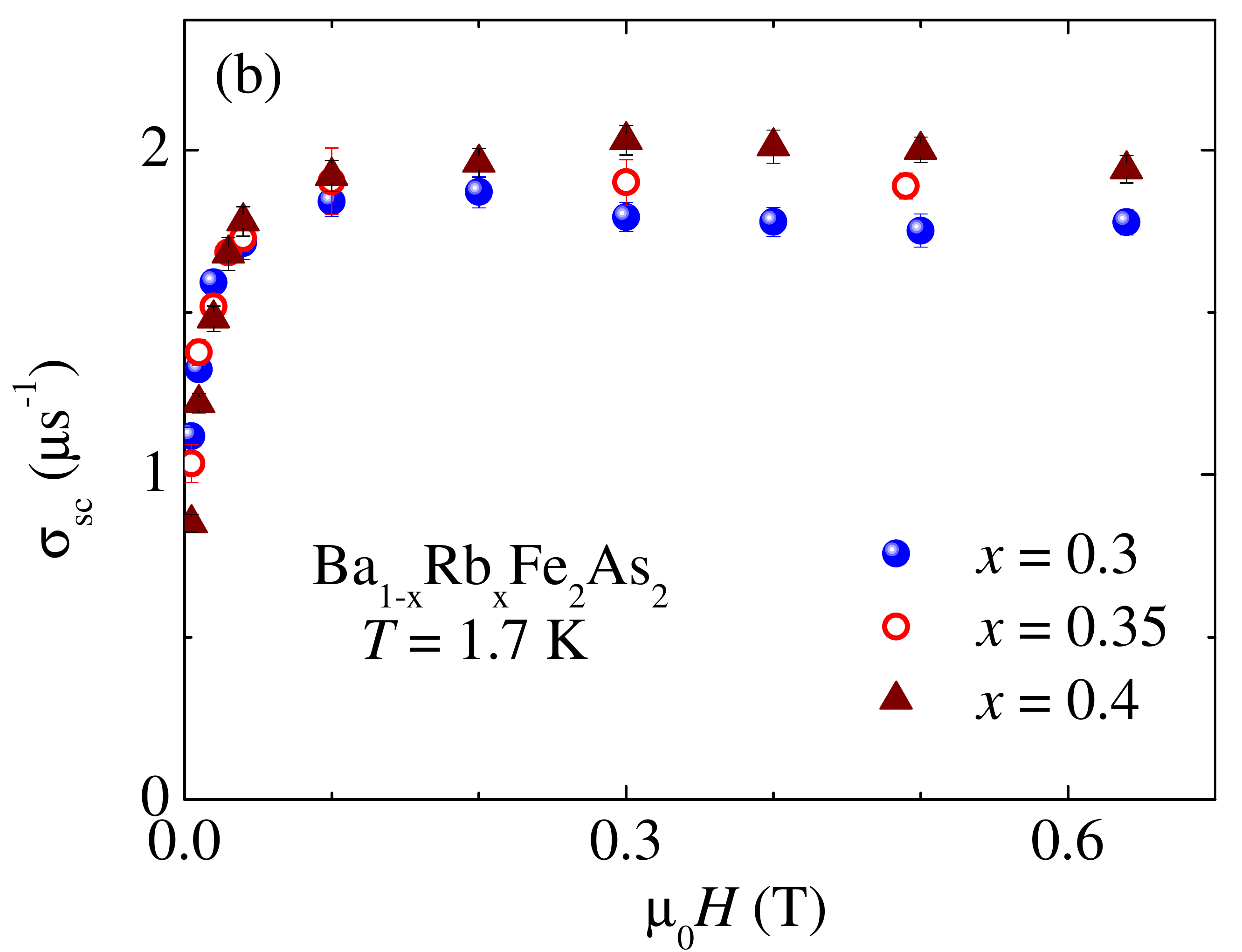}
\caption{ (Color online) (a) Temperature dependence of the superconducting muon spin depolarization rate
${\sigma}_{\rm sc}$ measured in an applied magnetic field of ${\mu}_{\rm 0}H = 0.04$~T for   
Ba$_{1-x}$Rb$_{x}$Fe$_{2}$As$_{2}$ ($x = 0.3$, 0.35, 0.4). (b) Field dependence of ${\sigma}_{\rm sc}$ 
at 1.7 K.}
\label{fig2}
\end{figure}

 It was reported \cite{khasanovFInd,Williams,Sonier2011} that in some iron-based
superconductors BaFe$_{2-x}$Co$_{x}$As$_{2}$ and SrFe$_{2-x}$Co$_{x}$As$_{2}$ field induced magnetism exists.
In the present work TF-${\mu}$SR spectra measured in a different applied fields 
(see Fig.~1 for ${\mu}_{\rm 0}H=0.04$~T) exhibit a Gaussian-like depolarization
above and below $T_{\rm c}$ that is typical of nuclear moments and the vortex lattice in polycrystalline samples,
respectively. In the presence of dilute or fast fluctuating
electronic moments one expects an exponential depolarization of TF-${\mu}$SR
spectrum, which is absent in the present case. 
Moreover, the SC muon depolarization rate ${\sigma}_{\rm sc}$ is constant at high fields as shown in Fig.~\ref{fig2}b.
In addition we observed a diamagnetic shift of the internal magnetic field ${\mu}_{\rm 0}$$H_{\rm int}$
sensed by the muons below $T_{\rm c}$. This is evident in Fig.~2b, where we plot the
difference between the internal field ${\mu}_{\rm 0}$$H_{\rm int,SC}$ measured in SC state 
and one ${\mu}_{\rm 0}$$H_{\rm int,NS}$ measured in the normal state at 
$T$ = 42 K.
Note, that in the systems BaFe$_{2-x}$Co$_{x}$As$_{2}$ and SrFe$_{2-x}$Co$_{x}$As$_{2}$, where
the field induced magnetism was detected, paramagnetic shift was observed \cite{khasanovFInd,Williams,Sonier2011} 
instead of the expected diamagnetic shift imposed by the SC state. 
All these observations indicate that there is no field induced magnetism in the system
Ba$_{1-x}$Rb$_{x}$Fe$_{2}$As$_{2}$ down to 1.7 K. The absence of magnetism in Ba$_{1-x}$Rb$_{x}$Fe$_{2}$As$_{2}$ implies that
the increase of the TF relaxation rate below
$T_{\rm c}$ is attributed entirely to the vortex lattice.

\begin{figure*}[ht!]
\centering
\includegraphics[width=1\linewidth]{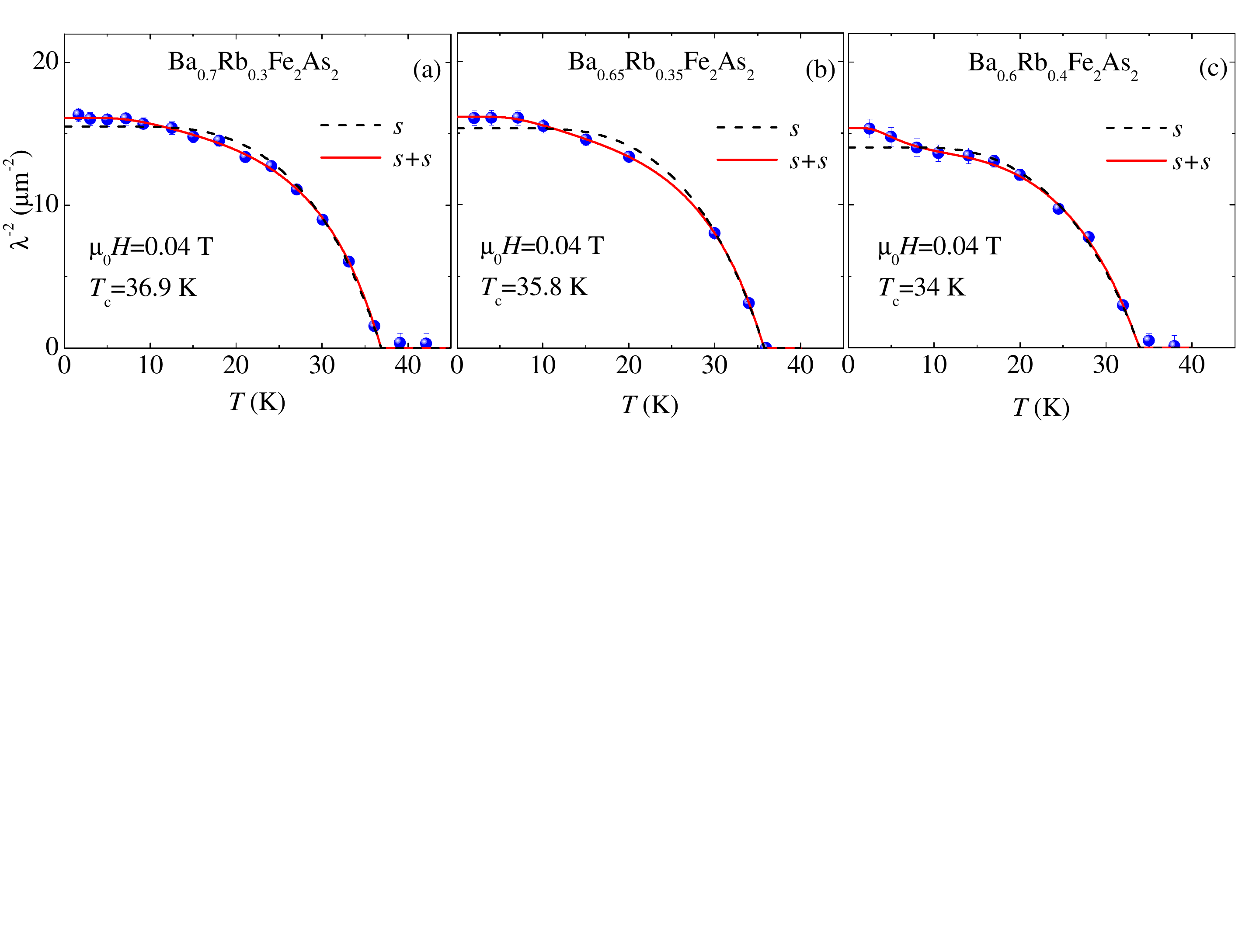}
\vspace{-7.8cm}
\caption{ (Color online) The temperature dependence of ${\lambda}^{-2}$ for
Ba$_{1-x}$Rb$_{x}$Fe$_{2}$As$_{2}$, measured in an applied field of
${\mu}_{\rm 0}H=0.04$~T: (a) $x=0.3$, (b) $x=0.35$ and (c) $x=0.4$. The dashed lines correspond
to a single gap BCS $s$-wave model, whereas the solid ones represent a fit using a two-gap 
($s+s$)-wave model.}
\label{fig3}
\end{figure*}
  
 The TF ${\mu}$SR data were analyzed by using the following functional form:\cite{Bastian}
\begin{equation}
P(t)=A\exp\Big[-\frac{(\sigma_{sc}^2+\sigma_{nm}^2)t^2}{2}\Big]\cos(\gamma_{\mu}B_{int}t+\varphi), 
\label{eq1}
\end{equation}

 Here $A$ denotes the initial assymmetry, $\gamma/(2{\pi})\simeq 135.5$~MHz/T 
is the muon gyromagnetic ratio, and ${\varphi}$ is the initial phase of the muon-spin ensemble. $B_{\rm int}$ represents the
internal magnetic field at the muon site, and the relaxation rates ${\sigma}_{\rm sc}$ 
and ${\sigma}_{\rm nm}$ characterize the damping due to the formation of the FLL in the superconducting state and of the nuclear 
magnetic dipolar contribution, respectively. In the analysis ${\sigma}_{\rm nm}$ was assumed 
to be constant over the entire temperature range and was fixed to the value obtained above 
$T_{\rm c}$ where only nuclear magnetic moments contribute to the muon depolarization rate ${\sigma}$.
As indicated by the solid lines in Fig.~\ref{fig1}, the ${\mu}$SR data are well described by Eq.~(1).
The temperature dependence of ${\sigma}_{\rm sc}$ for Ba$_{1-x}$Rb$_{x}$Fe$_{2}$As$_{2}$  (x = 0.3, 0.35, and 0.4) 
at ${\mu}_{\rm 0}H=0.04$~T is shown in Fig.~\ref{fig2}a.
Below $T_{\rm c}$ the relaxation rate ${\sigma}_{\rm sc}$ starts to increase from zero due to the
formation of the FLL. 
 
 For polycrystalline samples the temperature dependence of the London magnetic penetration depth 
${\lambda}(T)$ is related to the superconducting part of the Gaussian muon spin 
depolarization rate ${\sigma}_{\rm sc}(T)$ by the equation:\cite{Brandt}
\begin{equation}
\frac{\sigma_{sc}^2(T)}{\gamma_\mu^2}=0.00371\frac{\Phi_0^2}{\lambda^4(T)},
\end{equation}          
where ${\Phi}_{\rm 0}=2.068 {\times} 10^{-15}$~Wb is the magnetic-flux quantum. 
Equation (2) is only valid, when
the separation between the vortices is smaller than ${\lambda}$. In this case according
to the London model ${\sigma}_{\rm sc}$ is field independent.\cite{Brandt} We measured ${\sigma}_{\rm sc}$
as a function of the applied field at 1.7~K (see Fig.~\ref{fig2}b). Each point was obtained 
by field cooling the sample from above $T_{\rm c}$ to 1.7~K. First
${\sigma}_{\rm sc}$ strongly increases with increasing magnetic field until reaching a maximum
at ${\mu}_{\rm 0}H$  ${\simeq}$  0.03~T and then above 0.03~T stays nearly constant up to the highest field 
(0.64~T) investigated. Such a behavior is expected within the London model and is typical for polycrystalline
high temperature superconductors (HTS's).\cite{Pumpin} The observed field dependence of ${\sigma}_{\rm sc}$ implies that for a reliable determination of
the penetration depth the applied field must be larger than ${\mu}_{\rm 0}H= 0.03$~T.
 
 ${\lambda}$($T$) can be calculated within the local (London) approximation (${\lambda}$ ${\gg}$ ${\xi}$) 
by the following expression:\cite{Bastian,Tinkham}
\begin{equation}
\frac{\lambda^{-2}(T,\Delta_{0,i})}{\lambda^{-2}(0,\Delta_{0,i})}=
1+\frac{1}{\pi}\int_{0}^{2\pi}\int_{\Delta(_{T,\varphi})}^{\infty}(\frac{\partial f}{\partial E})\frac{EdEd\varphi}{\sqrt{E^2-\Delta_i(T,\varphi)^2}},
\end{equation}
where $f=[1+\exp(E/k_{\rm B}T)]^{-1}$ is the Fermi function, ${\varphi}$ is the angle along the Fermi surface, and ${\Delta}_{i}(T,{\varphi})={\Delta}_{0,i}{\delta}(T/T_{\rm c})g({\varphi}$)
(${\Delta}_{0,i}$ is the maximum gap value at $T=0$). 
The temperature dependence of the gap is approximated by the expression 
${\delta}(T/T_{\rm c})=\tanh{\{}1.82[1.018(T_{\rm c}/T-1)]^{0.51}{\}}$,\cite{carrington} 
while $g({\varphi}$) describes 
the angular dependence of the gap and it is replaced by 1 for both an $s$-wave and an $s$+$s$-wave gap,
and ${\mid}\cos(2{\varphi}){\mid}$ for a $d$-wave gap.\cite{Fang}   

 The temperature dependence of the penetration depth was analyzed using either a single gap or a two-gap model 
which is based on the so-called ${\alpha}$ model. This model was first discussed by 
Padamsee $et$ $al$.\cite{padamsee} 
and later on was succesfully used to analyse the magnetic penetration depth data in HTS's.\cite{carrington,khasanovalpha}
According to the ${\alpha}$ model, the superfluid density is calculated for each component using Eq.~3
and then the contributions from the two components added together, $i.e.$,  
\begin{equation}
\frac{\lambda^{-2}(T)}{\lambda^{-2}(0)}=\omega_1\frac{\lambda^{-2}(T,\Delta_{0,1})}{\lambda^{-2}(0,\Delta_{0,1})}+\omega_2\frac{\lambda^{-2}(T,\Delta_{0,2})}{\lambda^{-2}(0,\Delta_{0,2})},
\end{equation}\\
where ${\lambda^{-2}}(0)$ is the penetration depth at zero temperature, ${\Delta_{0,i}}$ is the 
value of the $i$th ($i=1$, 2) superconducting gap at $T=0$~K, and ${\omega}_{i}$ is a weighting 
factor which measures their relative contributions to ${\lambda^{-2}}$ (${\omega}_{1}+{\omega}_{2}=1$). 
\begin{table}[b!]
\caption{Summary of the parameters obtained for polycrystalline samples of  Ba$_{1-x}$Rb$_{x}$Fe$_{2}$As$_{2}$ ($x$ = 0.3, 0.35, 0.4, 1) by means of ${\mu}$SR. The data for $x$ = 1.0 are taken from Ref.~10.}
\vspace{0.3cm}
\begin{tabular}{lcccc}
\hline
\hline
                                               &  $x=0.3$ & $x=0.35$ & $x=0.4$ & $x=1.0$ \\ \hline 
$T_{\rm c}$ (K)                                &  36.9  & 35.8 & 34  & 2.52\\ 
${\Delta}_{1}$ (meV)                         &  3.2(7) & 2.9(8) & 1.1(3) & 0.15(2) \\
2${\Delta}_{1}/k_{\rm B}T_{\rm c}$       & 2.0(5) & 1.9(4) & 0.8(6) & 1.4(2)\\ 
${\Delta}_{2}$ (meV)                         &  9.2(3) & 8.8(3) & 7.5(2) & 0.49(4) \\ 
2${\Delta}_{2}/k_{\rm B}T_{\rm c}$       & 5.8(6) & 5.7(5) & 5.1(4) & 4.5(4)\\ 
${\omega}_{1}$                               & 0.19(5) & 0.21(4) & 0.15(3) & 0.36(3)\\
${\lambda}$ (nm)                       & 249(15) & 250 (17) & 255 (9) & 267(5) \\  \hline
\hline         
\end{tabular}
\label{table1}
\end{table}

 The results of the analysis for Ba$_{1-x}$Rb$_{x}$Fe$_{2}$As$_{2}$ ($x=0.3$, 0.35, 0.4) 
are presented in Fig.~\ref{fig3}. The dashed and the solid lines represent a fit to the data using
a $s$-wave and a $s+s$-wave models, respectively. The analysis appears to rule out the simple $s$-wave
model as an adequate description of ${\lambda}(T)$ for Ba$_{1-x}$Rb$_{x}$Fe$_{2}$As$_{2}$ ($x=0.3$, 0.35, 0.4).
A $d$-wave gap symmetry was also tested, but was found to be inconsistent with the data. The two-gap 
$s$+$s$-wave scenario with a small gap ${\Delta}_{1}$ and a large gap
${\Delta}_{2}$, describes the experimental data remarkably well.
The results of all samples extracted from the data analysis are  summarized in Table~\ref{table1}.
A two-gap scenario is in line with the generally accepted view of 
multi-gap superconductivity in Fe-based HTS.\cite{Evtushinsky,Ding,Khasanov,khasanov1,Bendele,Luetekns1} 
The magnitudes of the large and the small gap for 
Ba$_{1-x}$Rb$_{x}$Fe$_{2}$As$_{2}$ ($x=0.3$, 0.35, 0.4) (see Table~\ref{table1}) are in good agreement with the
results of a previous report.\cite{Evtushinsky} There it was pointed out that
most Fe-based HTS's exhibit two-gap superconducting behavior, characterized by a large gap with $2{\Delta}/k_{\rm B}T_{\rm c}=7(2)$ 
and a small one with 2.5(1.5). 
\begin{figure}[t!]
\centering
\includegraphics[width=1.6\linewidth]{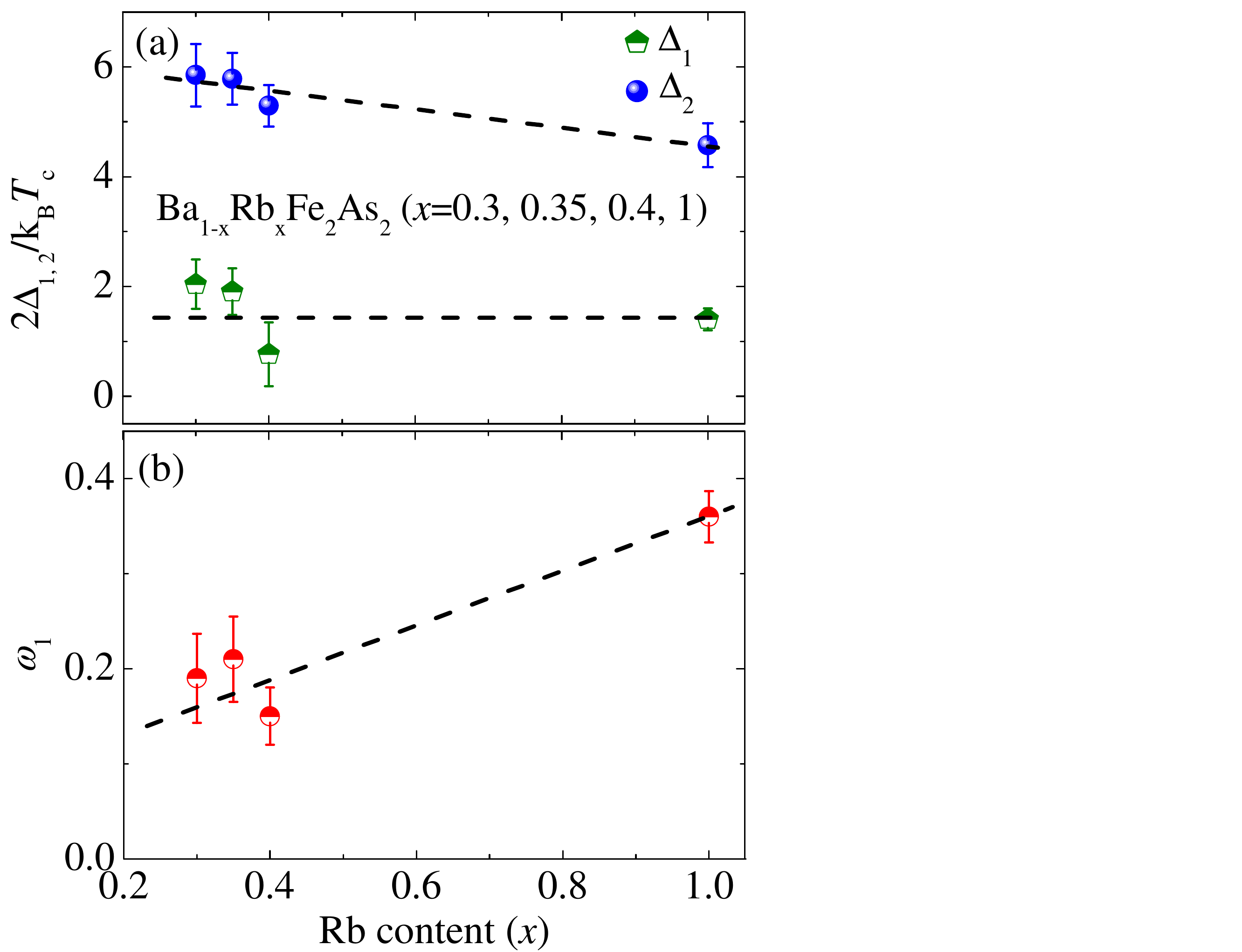}
\caption{ (Color online) Superconducting gap to $T_{c}$ ratios 2${\Delta}_{1,2}/k_{\rm B}T_{\rm c}$ (a)
and the contribution ${\omega}_{1}$ of the small gap to the superfluid density (b) 
as a function of the Rb composition for Ba$_{1-x}$Rb$_{x}$Fe$_{2}$As$_{2}$ ($x$ = 0.3, 0.35, 0.4, 1.0).
The measurements were performed in an applied magnetic field of ${\mu}_{\rm 0}H=0.04$~T. 
The data for RbFe$_{2}$As$_{2}$ are taken from Ref.~10. The dashed lines represent the
guides to the eyes.}
\label{fig4}
\end{figure}
In order to reach a more complete view of the superconducting properties
of Ba$_{1-x}$Rb$_{x}$Fe$_{2}$As$_{2}$ as a function of the Rb composition (hole-doping),
we combined the present data with the previous ${\mu}$SR results
on RbFe$_{2}$As$_{2}$ \cite{Shermadini} which presents the case of a naturally over-doped system.
Figure~\ref{fig4} shows the small gap to $T_{c}$ ratio $2{\Delta}_{1}/k_{\rm B}T_{\rm c}$, 
the large gap to $T_{c}$ ratio $2{\Delta}_{2}/k_{\rm B}T_{\rm c}$, and 
the weight ${\omega}_{1}$ of the small gap to the superfluid density as a function of Rb concentration.
The data for RbFe$_{2}$As$_{2}$ are taken from Ref.~\onlinecite{Shermadini}. Interestingly, the ratio $2{\Delta}_{2}/k_{\rm B}T_{\rm c}$
decreases with increasing $x$. On the other hand, the ratio $2{\Delta}_{1}/k_{\rm B}T_{\rm c}$
for the small gap is essentially independent of $x$. 
In addition, the weighting factor ${\omega}_{1}$ is found 
to increase with increasing $x$. We note that in the optimally doped 122-system Ba$_{1-x}$K$_{x}$Fe$_{2}$As$_{2}$ several bands cross the Fermi surface (FS).\cite{Evtushinsky,Ding,Zabolotnyy}
They consist of inner (${\alpha}$) and outer (${\beta}$) hole-like bands, 
both centered at the zone center ${\Gamma}$, and an electron-like band 
(${\gamma}$) centered at the M point. The superconducting gap opened
on the ${\beta}$ band was found to be smaller than those on the ${\alpha}$ and 
${\gamma}$ bands. 
It was proposed that the enhanced interband scattering between the   
${\alpha}$ and ${\gamma}$ bands might promote the kinetic process of pair scattering 
between these two FSs, leading to an increase of the pairing amplitude.\cite{Sato}
Hole doping may cause a shift of the band bottom of the electron pockets
above the Fermi level $E_{F}$. As a result, the interband scattering between ${\alpha}$ and ${\gamma}$
bands would diminish, since the ${\gamma}$ band is in the unoccupied side and
concomitantly the size of the ${\alpha}$ band is increased. According to ARPES results,
a decrease of interband scattering will lead to a decrease of pairing amplitude and the ratio $2{\Delta}/k_{\rm B}T_{\rm c}$
in agreement with the results presented in Fig.~\ref{fig4}a. These results suggest 
the possible role of interband processes in optimally hole-doped iron-based 122 superconductors.\cite{Ding,Sato}  
 
\begin{figure}[t!]
\centering
\includegraphics[width=1\linewidth]{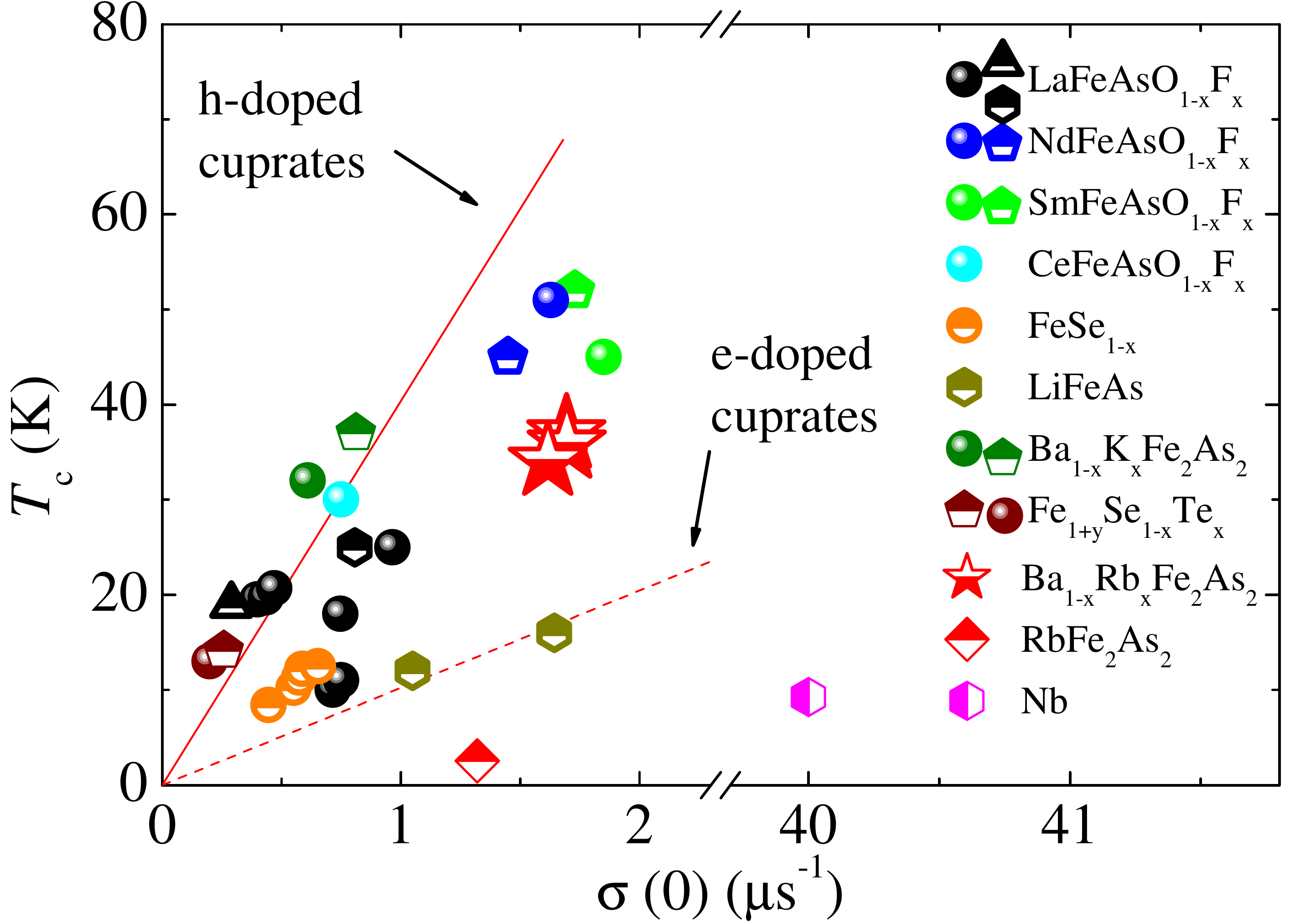}
\vspace{-0.5cm}
\caption{(Color online) Uemura plot for hole and electron doped
high $T_{\rm c}$ Fe-based superconductors (after Ref.~\onlinecite{Bendele}). The Uemura relation 
observed for underdoped cuprates is also shown (solid line for hole doping
and dashed line for electron doping) (after Ref.~\onlinecite{Shengelaya}). The point for conventional
BCS superconductor Nb is also shown. Data points for the pnictides are taken
from Refs.(\onlinecite{khasanov1,Bendele,Luetekns1},\onlinecite{Luetkens,Kim,Takeshita,Carlo,Khasanov2,Khasanov3,Pratt}). The stars show the data 
for Ba$_{1-x}$Rb$_{x}$Fe$_{2}$As$_{2}$ ($x$ = 0.3, 0.35, 0.4) obtained in this work. 
The point for RbFe$_{2}$As$_{2}$ is taken from Ref.~\onlinecite{Shermadini}.}
\label{fig5}
\end{figure}
One of the most interesting results of ${\mu}$SR investigations in HTS's is the observation of a remarkable
proportionality between $T_{\rm c}$ and the zero-temperature relaxation rate 
${\sigma}$(0)~${\propto}$~1/${\lambda}^{2}$(0)
(Uemura relation).\cite{Uemura1}
This relation $T_{\rm c}$(${\sigma}$) which 
seems to be generic for various families of cuprate HTS's, has the features that upon increasing the charge carrier doping 
$T_{\rm c}$ first increases linearly in the under-doped region (Uemura line), then saturates, and finally is suppressed for 
high carrier doping. The initial linear trend of the Uemura relation indicates that for these unconventional HTS's the ratio $T_c/E_F$ ($E_{\rm F}$ is the Fermi energy) is up to two orders of magnitude larger than for conventional BCS superconductors.         
Figure~\ref{fig5} shows $T_{\rm c}$ vs ${\sigma}(0)$ plot for various hole- and electron-doped high 
$T_{\rm c}$ Fe-based superconductors (after Ref.~\onlinecite{Bendele}), including the present results.
Solid line shows the Uemura relation in hole-doped cuprates \cite{Uemura1}
and dashed line corresponds to electron-doped cuprates as observed by Shengelaya 
$et$ $al$.\cite{Shengelaya}
The Uemura relation for Fe-based superconductors was already discussed in Ref.~\onlinecite{Bendele}. 
Here, we demonstrate that the data points for Ba$_{1-x}$Rb$_{x}$Fe$_{2}$As$_{2}$ ($x$ = 0.3, 0.35, 0.4) 
are located in the Uemura plot close to those of the other Fe-based superconductors. 
On the other hand, for naturally fully overdoped RbFe$_2$As$_2$, the ratio $T_c/\sigma(0)$ is strongly reduced.
The small value of ratio $T_c/\sigma(0)$ is characteristic for conventional superconductors.  
For comparison the point for conventional BCS superconductor Nb is also shown on the Uemura plot.
This suggests that superconductivity in the compound RbFe$_2$As$_2$ has more conventional character.
Additional experiments are in progress to clarify this point.

\section{SUMMARY AND CONCLUSIONS}

 In summary, we performed transverse-field ${\mu}$SR measurements of the
magnetic penetration depth ${\lambda}$ on polycrystalline samples of the
iron-based HTS's Ba$_{1-x}$Rb$_{x}$Fe$_{2}$As$_{2}$ ($x=0.3$, 0.35, 0.4).
The values of the superconducting transition temperature $T_{\rm c}$
and the zero temperature values of ${\lambda}$ were estimated to be  
$T_{\rm c}=36.9$~K, 35.8 K, 34 K and ${\lambda}(0)=249(15)$~nm, 250(17) nm, 255(9) nm 
for $x=0.3$, 0.35 and 0.4, respectively.
The temperature dependence of ${\lambda}$ is well described by a two-gap
$s$+$s$-wave scenario with gap values similar to 
Ba$_{1-x}$K$_{x}$Fe$_{2}$As$_{2}$.\cite{Khasanov,Evtushinsky}
ARPES investigations of 
Ba$_{1-x}$K$_{x}$Fe$_{2}$As$_{2}$ revealed that the large gap opens on the inner hole-like 
Fermi surface (${\alpha}$-band) centered at the ${\Gamma}$ point and 
on the electron-like FS (${\gamma}$-band) centered at the M point (tetragonal structure notations), while
the small gap opens on the outer hole-like band (${\beta}$) of the ${\Gamma}$ point.\cite{Sato}
We found that the large gap to $T_{\rm c}$ ratio  $2{\Delta}_{2}/k_{\rm B}T_{\rm c}$
decreases with increasing Rb content $x$. On the other hand, for the small
gap opening on the ${\alpha}$ and ${\gamma}$ bands, the ratio $2{\Delta}_{1}/k_{\rm B}T_{\rm c}$
is practically independent of $x$. In addition, the contribution of the small
gap ${\omega}_{1}$ to the total superfluid density  
increases with increasing $x$. These results may be interpreted by assuming
a disappearance of the electron pocket from the Fermi surface upon the high hole doping, resulting in a suppression of the scattering processes between the ${\alpha}$ and ${\gamma}$ bands. This
might cause the reduction of $T_{\rm c}$ for the overdoped RbFe$_{2}$As$_{2}$.
We also performed zero-field ${\mu}$SR experiments and found no evidence of either static 
or fluctuating magnetism, implying that the spin-density wave ordering of the Fe moments is completely
suppressed upon Rb doping. The absence of field induced magnetism in the investigated
compounds is also demonstrated. 
Finally, the correlation between $T_{\rm c}$ and the zero-temperature relaxation rate 
${\sigma}$(0) ${\propto}$ 1/${\lambda}^{2}$(0) is discussed for the samples
Ba$_{1-x}$Rb$_{x}$Fe$_{2}$As$_{2}$ ($x$ = 0.3, 0.35, 0.4, 1) using the Uemura classification scheme.

\section{Acknowledgments}~
~Part of this work was performed at the Swiss Muon Source, Paul Scherrer Institut,
Villigen, Switzerland. This work was supported by the Swiss National Science Foundation, the
SCOPES grant No. IZ73Z0${\_}$128242, the NCCR Project MaNEP, the EU Project
CoMePhS, and the Georgian National Science Foundation grant
GNSF/ST08/4-416.

\end{document}